\documentclass[useAMS]{article}
\usepackage{bookstyle}
\usepackage{graphicx}
\usepackage{cite}
\usepackage{amsfonts}
\begin{document}

\title{Dark Matter on the Smallest Scales}

\author{E. R. Siegel$^1$, G. D'Amico$^2$, E. Di Napoli$^3$,
L. Fu$^4$, M. P. Hertzberg$^5$, N. T. T. Huong$^6$, F.
Palorini$^7$, A. Sellerholm$^{8}$}

\address{$^1$Department of Physics, University of Wisconsin, 1150 University Avenue, Madison, WI 53706, USA\\
$^2$Department of Physics, Scuola Normale Superiore, Pisa 56100,
Italy\\
$^3$Department of Physics and Astronomy, University of North
Carolina, CB{\#}3255 Phillips Hall, Chapel Hill, NC 27599, USA\\
$^4$Joint Center of Astrophysics, Shanghai Normal University,
200234 Shanghai, China\\
$^5$Center for Theoretical Physics, MIT, Cambridge, MA 02139, USA\\
%%$^6$Institute of Physics, Federal University of Rio de Janeiro,
%%Rio de Janeiro, RJ 21941-972, Brazil\\
%$^6$D\`{e}partement de Physique Th\`{e}orique, Universit\`{e}
%Libre de Bruxelles, 1050 Bruxelles, Belgium\\
$^6$Centre for High Energy Physics, Vietnam National University -
Hanoi, 334 Nguyen Trai Street, Hanoi,
Vietnam\\
$^7$Universit\`{e} C. Bernard, 69662 Villeurbanne cedex, France\\
$^{8}$Department of Physics, Stockholm University, 10691
Stockholm, Sweden}

\photo{\includegraphics{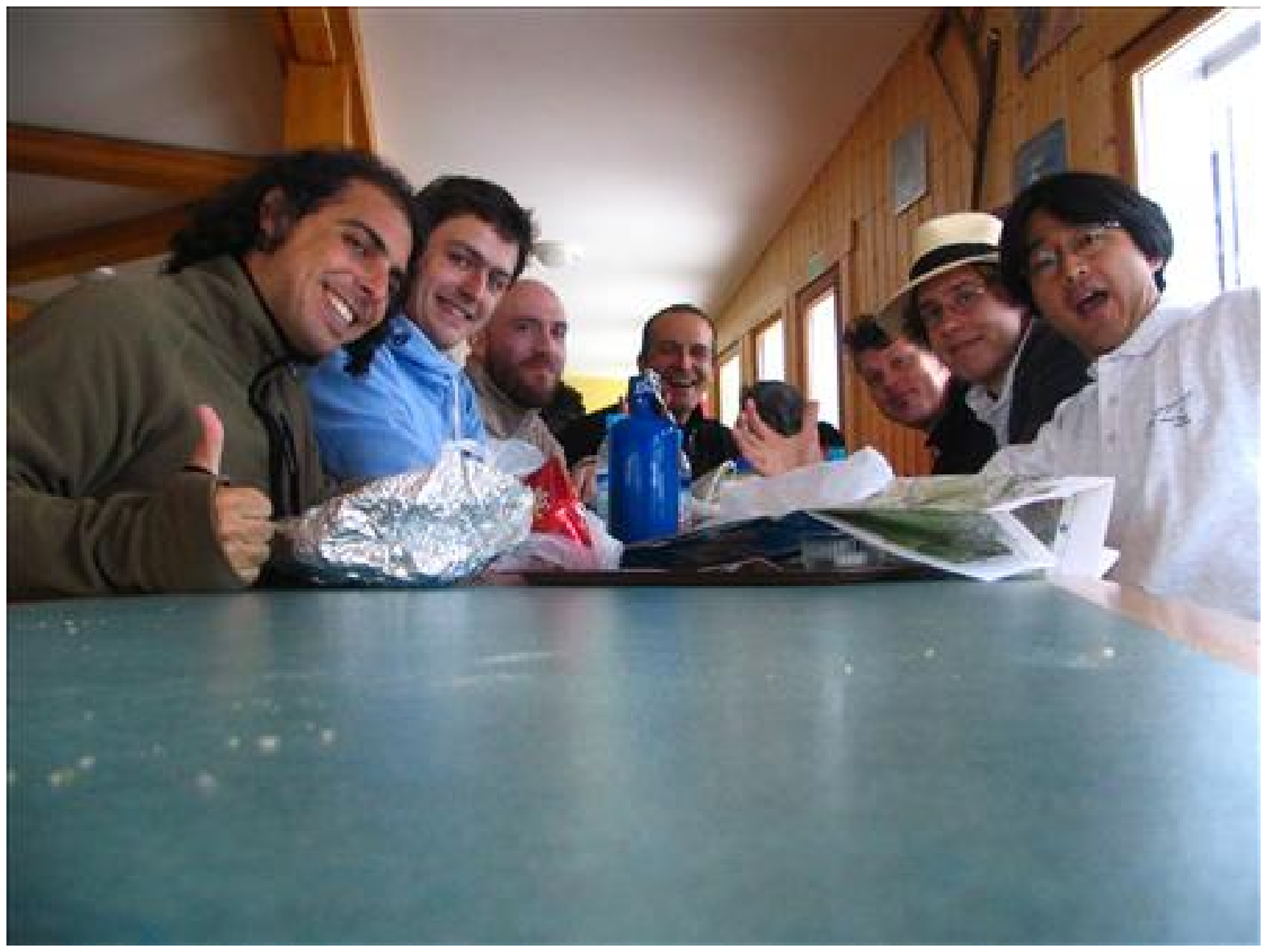}}

\frontmatter \maketitle \mainmatter
%
%This would have been the abstract:
%
\abstract{This work investigates the dark matters structures that
form on the smallest cosmological scales.  We find that the types
and abundances of structures which form at approximately
Earth-mass scales are very sensitive to the nature of dark matter.
We explore various candidates for dark matter and determine the
corresponding properties of small-scale structure.  In particular,
we discuss possibilities for indirect detection of dark matter
through small-scale structure, and comment on the potential of
these methods for discriminating between dark matter candidates.}

\section{Introduction \label{intro}}

Perhaps the most fundamental goal of cosmological physics is to
determine the composition and evolution of the universe. Cosmology
has recently reached a state where the vast majority of
measurements point towards the same result; the universe is
composed of $\sim 5$ per cent baryons, $\sim 20$ per cent dark
matter, and $\sim 75$ per cent dark energy.  This result is quite
puzzling, as dark matter and dark energy have only been detected
by their gravitational effects on cosmological scales.  While much
is known about baryonic structure as well as the particle
properties of baryons, much less is known about dark matter
structure and its particle properties, and even less is known
about dark energy.  As a result, the nature of these two latter
components of the universe, which compose $95$ per cent of its
total energy density, are presently the greatest unsolved puzzles
in cosmology.

This work investigates the nature of dark matter by examining its
effects on cosmologically small scales.  In particular,
constraining the properties of small-scale structure will
constrain the nature of dark matter.  The layout of this paper is
as follows: section \ref{problem} presents an overview of the dark
matter problem, focusing on astrophysical evidence.  Section
\ref{candidates} identifies the leading particle candidates for
dark matter, with a view towards their distinguishing properties.
Section \ref{sss} describes and contrasts the properties of small
scale structure which forms in various dark matter models.
Finally, section \ref{detect} provides a discussion on possible
observable signals which could point towards the nature of dark
matter.

\section{The Dark Matter Problem \label{problem}}

The discrepancy between the amount of mass inferred from
observations of light and the amount of mass inferred from
gravitation on cosmological scales is known as the dark matter
problem.  In our solar system, nearly all of the mass is contained
in the sun, and so it seems reasonable to assume that light traces
mass.  Since properties of stars are well known, it is possible to
infer the amount of mass present in luminous sources $(M_\star)$,
such as stars. From measurements of $M_\star$, it is possible to
infer the fractional energy density in stars, $\Omega_\star$,
where $\Omega_i$ is defined as the energy density in species $i$
over the critical energy density, $\rho_{c}$.  Observations
indicate that $\Omega_\star \simeq 0.005$ \cite{FP:04}, or that
luminous matter composes about $0.5$ per cent of the total energy
density in the universe.

On the other hand, it is also possible to reconstruct the matter
density of the universe from its gravitational mass (yielding
$\Omega_m$) instead of its luminous mass ($\Omega_\star$).  Many
methods exist for determining $\Omega_m$, including extrapolating
the matter density from peculiar velocities within galaxy clusters
\cite{Zwicky}, rotational and virial properties of individual
galaxies \cite{rotate}, gravitational lensing \cite{lensing}, and
from large scale structure \cite{2dF}.  All of these methods yield
a range that $0.15 < \Omega_m < 0.35$, consistent with one another
at the 2-$\sigma$ level.  The best measurement of matter density
comes from a combination of the latest microwave background data
\cite{Spergel:06} with type Ia supernova data \cite{Riess:06},
yielding $\Omega_m = 0.26 \pm 0.02$.  This discrepancy between
$\Omega_m$ and $\Omega_\star$ is the dark matter problem; the
amount of mass in stars is only $2$ per cent of the total matter
density!  The remaining $98$ per cent, the non-luminous matter, is
known as dark matter.

\section{Dark Matter Candidates \label{candidates}}

Once the existence of dark matter has been established, it becomes
a fundamental question to inquire what its nature is.  Many
candidates have been proposed (see \cite{BHS} for a review), but a
confirmed experimental detection has been thus far elusive.  This
section therefore investigates the properties and viability of
four generic models of dark matter: baryons, neutrinos, thermal
relics, and nonthermal relics.

Baryons are the most obvious candidate for dark matter, as
baryonic dark matter is observed in the form of planets, low
luminosity stars, and diffuse gas clouds.  One consequence is that
baryonic dark matter will fragment and collapse to form MAssive
Compact Halo Objects (MACHOs).  Unfortunately, the quantitative
effect of Silk damping in large-scale structure \cite{SilkLSS},
the insufficient number of MACHOs \cite{EROS}, and big bang
nucleosynthesis \cite{bbn} all indicate that $\Omega_m \simeq
0.04$.  Finally, direct observations of X-rays from interacting
galaxies show that dark matter does not interact in the same
fashion as baryons \cite{bullet}, which means that although
baryons compose a significant fraction (about $15$ per cent) of
the total matter, they cannot be responsible for all of the dark
matter.

It is therefore unavoidable that the majority of dark matter be
non-baryonic.  The standard model contains a candidate for
non-baryonic dark matter: the neutrino.  However, neutrinos are
produced thermally and have very low masses, conflicting with
constraints from the Lyman-$\alpha$ forest that limit the mass of
thermal dark matter to be at least $2 \, \mathrm{keV}$ \cite{Lya}.
These observations, combined with the constraints on neutrino
masses, rule out standard model neutrinos as a significant
component of dark matter.

Heavier thermal relics are allowed experimentally, however, and
particle physics beyond the standard model provides us with many
well-motivated dark matter candidates of this type, including the
lightest supersymmetric particle and the lightest Kaluza-Klein
particle.  These particles do not interact either strongly or
electromagnetically, and are therefore classified as Weakly
Interacting Massive Particles (WIMPs).  Thermal relics are defined
by the fact that, at some point in the past history of the
universe, the particles were in thermal equilibrium with the
primordial plasma. At some point, their abundance freezes out (a
process known as chemical decoupling), and the number of dark
matter particles remains constant thereafter.  At a later point,
the dark matter particles cease to scatter off of the plasma
(kinetic decoupling), and thereafter evolve solely
gravitationally.

Nonthermal relics, on the other hand, were never in thermal
equilibrium with the primordial plasma.  Instead, these dark
matter candidate particles can be produced during phase
transitions in the early universe.  Candidates include axions
\cite{PQ} and massive gravitons \cite{Tinyakov}.  Unlike thermal
relics, nonthermal relics evolve only gravitationally from the
moment of their creation, receiving only a gravitational imprint
from the primordial plasma. As will be discussed in sections
\ref{sss} and \ref{detect}, present-day gravitational effects of
dark matter have the potential to shed light on its nature, and
may be able to distinguish between thermal and nonthermal relics.

\section{Small Scale Structure \label{sss}}

Cosmological structure formation is the key process that allows an
understanding of the types and abundances of structures which will
form in the universe given a set of initial conditions.  While
large-scale structure is well understood, small-scale structure is
more of a challenge due to the uncertainty of the nature of dark
matter and the problems associated with gravitational collapse in
the deeply nonlinear regime.  The remainder of the work in this
paper represents the initial stages of an ambitious research
project to probe the nature of dark matter via astrophysical
observations on small (i.e., sub-galactic) scales.

Just as baryonic structures are suppressed on mass scales below
$\sim 10^6$ solar masses $(M_\bigodot)$ due to Silk damping prior
to decoupling, WIMPy structures will be suppressed on scales
smaller than the horizon at the time of kinetic decoupling.  By
contrast, nonthermal relic dark matter will have no such
suppression, owing to the fact that they were always decoupled
from the primeval plasma.  The fluctuations at the time of
decoupling are frozen in, and the nonlinear structures on small
scales which eventually form from the collapse of dark matter will
be very sensitive to the epoch of kinetic decoupling. Hence, the
exact mass and abundance of these cosmic microstructures depend
very sensitively on the dark matter particle properties.  In
particular, they are sensitive to dark matter mass, which is
directly tied to the times of kinetic and chemical decoupling, and
to whether the dark matter was produced thermally or nonthermally.

One challenge of understanding small scale structure is to
understand the dark matter's evolution through kinetic decoupling.
This difficulty is evidenced by the varied results obtained by
authors using different approximations \cite{Green,Loeb}. We
assume that the dark matter is composed of neutralino WIMPs, and
follow the approach given by E. Bertschinger \cite{EdBert:06}.  By
calculating the transfer functions for cold dark matter
fluctuations beginning with the full Boltzmann equations
describing scattering between WIMPs and the plasma, the
uncertainty created by the aforementioned approximations is
eliminated.  We begin by treating the dark matter particles as an
unperturbed fluid.  Analysis allows us to obtain an expression for
the temperature at which kinetic decoupling occurs,
\begin{equation}
\label{decouple} T_d = 0.2528 \, g_{\mathrm{eff}}^{1/8} \left(
\frac{m_{\tilde{L}}^2-m_\chi^2}{ G_F m_W^2 m_\chi^2
\tan{\theta_W}} \right)^{1/2} \left(
\frac{m_\chi^5}{m_\mathrm{Pl}} \right)^{1/4} \mathrm{,}
\end{equation}
where $g_\mathrm{eff}$ is the number of thermal degrees of
freedom, $G_F$ is the Fermi constant, $\theta_W$ is the Weinberg
angle, and $m_\chi$, $m_{\tilde{L}}$, and $m_W$ are the masses of
the neutralino, slepton, and $W$ boson. There is an associated
time for kinetic decoupling, $t_d$. For a slepton mass of $200
\,$GeV, this yields a temperature for kinetic decoupling of
\begin{equation}
T_d \simeq 15.7 \left( \frac{m_\chi}{100 \, \mathrm{GeV}}
\right)^{5/4} \, \mathrm{MeV.}
\end{equation}
By including perturbations in the gravitational field, one can
derive the density transfer function for cold dark matter through
the epoch of kinetic decoupling.  We find that, on scales outside
the horizon at kinetic decoupling $(k / a < 1 / t_d)$, acoustic
oscillations average out, resulting in a logarithmic growth of
cold dark matter fluctuations.  However, on scales inside the
horizon $(k / a > 1 / t_d)$, density fluctuations exhibit damped
acoustic oscillations, which suppress the formation of structure.

The density transfer function can then be evolved through
electron-positron pair annihilation and matter-radiation equality,
yielding predictions for small scale cold dark matter structure.
The Press-Schechter mass fraction can be derived, and indicates a
suppression in the mass of collapsed structure per mass interval
on mass scales below $\approx 2.3 M_d$, where $M_d$ is the mass
contained in a typical density fluctuation at kinetic decoupling.
For a neutralino of mass $m_\chi = 100 \, \mathrm{GeV}$, this
indicates that the number of cold dark matter structures which
form on mass scales below approximately $20$ Earth masses will be
suppressed.  The root-mean-squared mass density perturbation
containing a mass $M$, $\sigma(M)$, is suppressed as
\begin{equation}
\frac{ d \sigma(M)}{d \ln{M}} \propto \left( \frac{M}{M_d}
\right)^{2/3}
\end{equation}
for masses $M \ll M_d$, which then results in a suppression of
WIMP microhalos of equivalent mass when converted into the
nonlinear regime. By comparison, nonthermal dark matter such as
axions does not have a suppression in its Press-Schechter mass
fraction, as it was always kinetically decoupled \cite{Zurek}. For
axion-like dark matter, this would indicate a comparatively much
larger number of collapsed structures below about $20$ Earth
masses, resulting in nonlinear structures we term Nonthermal
Axionic Collapsed HalOs (NACHOs).

\section{Detectability and Future Work \label{detect}}

It is unknown whether these dark matter microhalos, regardless of
whether they are WIMPs or NACHOs, will be able to survive
hierarchical mergers and galactic infall, and thus be found intact
within our own galaxy \cite{Diemand}.  Recent work
\cite{Gnedin:06}, however, indicates that this is a quantitative
question and not a qualitative one, as tidal stripping is not
sufficient to completely destroy all of these collapsed
structures. One extremely important question to ask is, "what does
small scale structure look like today?" $N-$body simulations yield
varied results, and there is no consensus as to the density
profiles and core concentrations of these objects.  Additionally,
the Press-Schechter approach may be invalid, as monolithic
collapse may be more important than hierarchical mergers on small
scales.  Ideally, therefore, detection methods which probe the
density profiles and abundances of these microhalos will be able
to not only constrain the nature of dark matter, but will provide
information about the types of structures which form through the
deeply nonlinear regime.

If these microhalos survive intact to the present day, a large
number should be present within our own galaxy.  We therefore seek
to uncover methods to detect these microhalos and investigate
their properties.  For neutralino dark matter, there will be a
significant annihilation cross-section, which could result in an
observable gamma-ray signal.  For a $100 \,$GeV WIMP forming an
Earth mass microhalo with an NFW profile, we calculate a gamma-ray
flux of $\sim 10^{22} \, \mathrm{photons} \, \,
\mathrm{sec}^{-1}$. As abundance estimates indicate that the
nearest microhalo should be nearer than the nearest star
\cite{Diemand}, this flux has the potential to outshine even
gamma-rays emitted from the galactic center.  Gravitational
lensing due to a microhalo transit is also a possible effect.
While WIMP microhalos are too diffuse, NACHOs may form much more
dense structures \cite{Zurek}, which may leave observable lensing
signals.  One very interesting possibility currently being
investigated by two of the authors is that a dark matter microhalo
transiting across the line-of-sight from an observer to a pulsar
could cause a shift in the pulse arrival time due to the
gravitational time delay \cite{Siegel:07}.  Finally, interactions
between dark matter microhalos and stars or gas clouds may be
important.  We note that if a dark matter microhalos of a few
Earth masses were gravitationally captured by our sun, it would
cause an anomalous acceleration towards the sun at significantly
large distances.

Cosmological structures on small scales have the potential to shed
light on the nature of dark matter, as many of the methods of
detecting small-scale dark matter structure are sensitive to its
mass and/or method of production.  Success of any of the above
methods would provide the first definitive confirmation of the
presence of dark matter within our own galaxy.  Although a
tremendous amount of nonlinear processing has occurred since the
creation of dark matter microhalos, present and future searches
for small-scale cosmological structures may hold the key to
determining the nature of the non-baryonic dark matter in our
universe.

\smallskip

We would like to thank the organizers of the 86th Les Houches
summer school, Francis Bernardeau and Christophe Grojean, as well
as the staff and our colleagues for providing a stimulating and
productive working environment.  We also acknowledge Sergio Joras,
Laura Lopez-Honorez and Gilles Vertongen for their contributions
to our group.  E.R.S. acknowledges Hai-Ping Cheng and Pierre
Ramond for support.


\begin{thebibliography}{1}

\bibitem{FP:04}
{\sc M.~Fukugita and P.~J.~E.~Peebles}, Astrophys.\ J.\  {\bf
616}, 643 (2004)

\bibitem{Zwicky}
{\sc R.~P.~van der Marel, J.~Magorrian, R.~G.~Carlberg,
H.~K.~C.~Yee and E.~Ellingson}, Astron.\ J.\ {\bf 119}, 2038
(2000)

\bibitem{rotate}
{\sc T.~G.~Brainerd and M.~A.~Specian}, Astrophys.\ J.\  {\bf
593}, L7 (2003)

\bibitem{lensing}
{\sc C.~Heymans {\it et al.}}, Mon.\ Not.\ Roy.\ Astron.\ Soc.\
Lett.\  {\bf 371}, L60 (2006)

\bibitem{2dF}
{\sc J.~A.~Peacock {\it et al.}}, Nature {\bf 410}, 169 (2001)

\bibitem{Spergel:06}
{\sc D.~N.~Spergel {\it et al.}}, arXiv:astro-ph/0603449.

\bibitem{Riess:06}
{\sc A.~G.~Riess {\it et al.}}, arXiv:astro-ph/0611572.

\bibitem{BHS}
{\sc G.~Bertone, D.~Hooper and J.~Silk}, Phys.\ Rept.\  {\bf 405},
279 (2005)

\bibitem{SilkLSS}
{\sc C.~J.~Miller, R.~C.~Nichol and X.~l.~Chen}, Astrophys.\ J.\
{\bf 579}, 483 (2002)

\bibitem{EROS}
{\sc C.~Alcock {\it et al.}  [MACHO Collaboration]}, Astrophys.\
J.\ {\bf 499}, L9 (1998)

\bibitem{bbn}
{\sc B.~D.~Fields and K.~A.~Olive}, Nuc.\ Phys.\ A {\bf 777}, 208
(2006)

\bibitem{bullet}
{\sc D.~Clowe, M.~Bradac, A.~H.~Gonzalez, M.~Markevitch,
S.~W.~Randall, C.~Jones and D.~Zaritsky}, arXiv:astro-ph/0608407.

\bibitem{Lya}
{\sc M.~Viel, J.~Lesgourgues, M.~G.~Haehnelt, S.~Matarrese and
A.~Riotto}, Phys.\ Rev.\ D {\bf 71}, 063534 (2005)

\bibitem{PQ}
{\sc R.~D.~Peccei and H.~R.~Quinn}, Phys.\ Rev.\ Lett.\  {\bf 38},
1440 (1977)

\bibitem{Tinyakov}
{\sc S.~L.~Dubovsky, P.~G.~Tinyakov and I.~I.~Tkachev}, Phys.\
Rev.\ Lett.\  {\bf 94}, 181102 (2005)

\bibitem{Green}
{\sc A.~M.~Green, S.~Hofmann and D.~J.~Schwarz}, JCAP {\bf 0508},
003 (2005)

\bibitem{Loeb}
{\sc A.~Loeb and M.~Zaldarriaga}, Phys.\ Rev.\ D {\bf 71}, 103520
(2005)

\bibitem{EdBert:06}
{\sc E.~Bertschinger}, Phys.\ Rev.\ D {\bf 74}, 063509 (2006)

\bibitem{Zurek}
{\sc K.~M.~Zurek, C.~J.~Hogan and T.~R.~Quinn},
arXiv:astro-ph/0607341.

\bibitem{Diemand}
{\sc J.~Diemand, B.~Moore and J.~Stadel}, Nature {\bf 433}, 389
(2005)

\bibitem{Gnedin:06}
{\sc T.~Goerdt, O.~Y.~Gnedin, B.~Moore, J.~Diemand and J.~Stadel},
arXiv:astro-ph/0608495.

\bibitem{Siegel:07}
{\sc E.~R.~Siegel, J.~N.~Fry and M.~P.~Hertzberg}, {\it in
preparation}.

\end{thebibliography}
\end{document}